\documentstyle[times]{ACMconf}

\title{Quantum vs.\ Classical Communication and Computation}

\author{
\parbox{2in}{
\begin{center}
Harry Buhrman\thanks{
Research  supported in part by the Dutch foundation for scientific
research (NWO) by SION project 612-34-002, and by the European Union through
NeuroCOLT ESPRIT Working Group Nr. 8556, and HC\&M grant nr. ERB4050PL93-0516.
Address: CWI, Quantum Computing and Advanced Systems Research,
P.O.\ Box 94079, Amsterdam, The Netherlands.
E-mail: {\tt buhrman@cwi.nl}.}\\CWI
\end{center}
}
\and
\parbox{2in}{
\begin{center}
Richard Cleve\thanks{
Research supported in part by Canada's  
NSERC. Address: Department of Computer Science, University of Calgary, 
Calgary, Alberta, Canada T2N 1N4.
E-mail: {\tt cleve@cpsc.ucalgary.ca}.}\\University of Calgary
\end{center}
}
\and
\parbox{2in}{
\begin{center}
Avi Wigderson\thanks{Work partially supported by grant 032-7736 from the 
Israel 
Academy of Sciences. Part of this work was done during a visit to the
Institute for Advanced Study, Princeton, supported by a Sloan
Foundation grant number 96-6-2.
Address: Hebrew University Jerusalem, Israel. E-mail: {\tt 
avi@cs.huji.ac.il}.}\\Hebrew University
\end{center}
}
}

\begin{document}


\maketitle
 
 
\newtheorem{THEOREM}{Theorem}[section]
\newenvironment{theorem}{\begin{THEOREM} \hspace{-.85em} {\bf :} 
}%
                        {\end{THEOREM}}
\newtheorem{LEMMA}[THEOREM]{Lemma}
\newenvironment{lemma}{\begin{LEMMA} \hspace{-.85em} {\bf :} }%
                      {\end{LEMMA}}
\newtheorem{FACT}[THEOREM]{Fact}
\newenvironment{fact}{\begin{FACT} \hspace{-.85em} {\bf :} }%
                      {\end{FACT}}
\newtheorem{COROLLARY}[THEOREM]{Corollary}
\newenvironment{corollary}{\begin{COROLLARY} \hspace{-.85em} {\bf 
:} }%
                          {\end{COROLLARY}}
\newtheorem{PROPOSITION}[THEOREM]{Proposition}
\newenvironment{proposition}{\begin{PROPOSITION} \hspace{-.85em} 
{\bf :} }%
                            {\end{PROPOSITION}}
\newtheorem{DEFINITION}[THEOREM]{Definition}
\newenvironment{definition}{\begin{DEFINITION} \hspace{-.85em} {\bf 
:} \rm}%
                            {\end{DEFINITION}}
\newtheorem{EXAMPLE}[THEOREM]{Example}
\newenvironment{example}{\begin{EXAMPLE} \hspace{-.85em} {\bf :} 
\rm}%
                            {\end{EXAMPLE}}
\newtheorem{CONJECTURE}[THEOREM]{Conjecture}
\newenvironment{conjecture}{\begin{CONJECTURE} \hspace{-.85em} 
{\bf :} \rm}%
                            {\end{CONJECTURE}}
\newtheorem{PROBLEM}[THEOREM]{Problem}
\newenvironment{problem}{\begin{PROBLEM} \hspace{-.85em} {\bf :} 
\rm}%
                            {\end{PROBLEM}}
\newtheorem{QUESTION}[THEOREM]{Question}
\newenvironment{question}{\begin{QUESTION} \hspace{-.85em} {\bf :} 
\rm}%
                            {\end{QUESTION}}
\newtheorem{REMARK}[THEOREM]{Remark}
\newenvironment{remark}{\begin{REMARK} \hspace{-.85em} {\bf :} 
\rm}%
                            {\end{REMARK}}
 
\newcommand{\thm}{\begin{theorem}}
\newcommand{\lem}{\begin{lemma}}
\newcommand{\pro}{\begin{proposition}}
\newcommand{\dfn}{\begin{definition}}
\newcommand{\rem}{\begin{remark}}
\newcommand{\xam}{\begin{example}}
\newcommand{\cnj}{\begin{conjecture}}
\newcommand{\prb}{\begin{problem}}
\newcommand{\que}{\begin{question}}
\newcommand{\cor}{\begin{corollary}}
\newcommand{\prf}{\noindent{\bf Proof:} }
\newcommand{\ethm}{\end{theorem}}
\newcommand{\elem}{\end{lemma}}
\newcommand{\epro}{\end{proposition}}
\newcommand{\edfn}{\bbox\end{definition}}
\newcommand{\erem}{\bbox\end{remark}}
\newcommand{\exam}{\bbox\end{example}}
\newcommand{\ecnj}{\bbox\end{conjecture}}
\newcommand{\eprb}{\bbox\end{problem}}
\newcommand{\eque}{\bbox\end{question}}
\newcommand{\ecor}{\end{corollary}}
\newcommand{\eprf}{\bbox}
\newcommand{\beqn}{\begin{equation}}
\newcommand{\eeqn}{\end{equation}}
\newcommand{\wbox}{\mbox{$\sqcap$\llap{$\sqcup$}}}
\newcommand{\bbox}{\vrule height7pt width4pt depth1pt}
\newcommand{\qed}{\bbox}
\def\sup{^}
\def\ra{\rightarrow}
\def\zok{\{0,1\}^k}
\def\zo{\{0,1\}}
\def\zom{\{0,1\}^m}
\def\zon{\{0,1\}^n}
\def\zor{\{0,1\}^r}
\def\zol{\{0,1\}^l}
\def\Tp{Tchebyshef polynomial}
\def\Tps{TchebysDeto be the maximafine $A(n,d)$ l size of a code with distance 
$d$hef polynomials}

\newcommand{\rarrow}{\rightarrow}
\newcommand{\larrow}{\leftarrow}

%
\def\half{\textstyle{1 \over 2}}
\def\sqhalf{\textstyle{1 \over \sqrt{2}}}
\def\01{\{0,1\}}
\def\x{\times}
\def\ox{\otimes}
\def\xor{+}
\def\AND{}
\def\Q{\mbox{\sf Q}}
\def\e{\varepsilon}
\def\ket#1{\mbox{$| #1 \rangle$}}
\def\sqm#1{\textstyle{\sqrt{#1}}}
\def\half{\textstyle{1 \over 2}}
\def\one{\mbox{\boldmath $1$}}
\def\nl{\newline}
\def\ni{\noindent}
\def\ii{\hspace*{8mm}}
\def\ee{\vspace*{1mm}}
\def\eee{\vspace*{5mm}}
\def\F{{\cal F}}
\def\H{{\cal H}}
\def\loud#1{\noindent{\bf #1 }}
\def\EQ{\mbox{\it EQ\/}}
\def\EQprime{\mbox{\it EQ\/}^{\prime}}
\def\IP{\mbox{\it IP\/}}
\def\DISJ{\mbox{\it DISJ\/}}
\def\PARITY{\mbox{\it PARITY\/}}
\def\BAL{\mbox{\it BAL\/}}
\def\STAB{\mbox{\it STAB\/}}
\def\OR{\mbox{\it OR\/}}
\def\AND{\mbox{\it AND\/}}
\def\MAJ{\mbox{\it MAJORITY\/}}
\def\MODq{\mbox{\it MOD}_q}
\def\SIGMA{\mbox{\it SIGMA}}
\def\PI{\mbox{\it PI}}

\overfullrule=0pt
\def\setof#1{\lbrace #1 \rbrace}
\newcommand{\card}[1]{{\mathopen{|\!|}#1\mathclose{|\!|}}}
\newcommand{\pair}[1]{{\mathopen<#1\mathclose>}}
\section*{Abstract}
We present a simple and general simulation technique that transforms any
black-box quantum algorithm ({\it \`a la} Grover's database search algorithm)
to a quantum communication protocol for a related problem, in a way
that fully exploits the quantum parallelism. This allows us to obtain
new positive and negative results.

The positive results are novel  quantum communication protocols that 
are built from nontrivial quantum algorithms via this simulation.
These protocols, combined with (old and new) classical lower bounds, 
are shown to provide the first asymptotic separation results between
the quantum and classical (probabilistic) two-party communication 
complexity models.
In particular, we obtain a quadratic separation for the bounded-error 
model, and an exponential separation for the zero-error model.

The negative results transform  known quantum communication lower 
bounds to computational lower bounds in the black-box model.
In particular, we show that the quadratic speed-up achieved by Grover 
for the $\OR$ function is impossible for the $\PARITY$ function or the $\MAJ$ 
function in the bounded-error model, nor is it possible for the $\OR$
function itself in the exact case.
This dichotomy naturally suggests a study of bounded-depth predicates 
(i.e.\ those in the polynomial hierarchy) between $\OR$ and $\MAJ$.
We present black-box algorithms that achieve near quadratic speed up for
all such predicates.


\section{Introduction and summary of results}

We discuss our results about quantum communication complexity and 
quantum black-box algorithms in separate subsections.
Regarding quantum communication complexity, Subsection 1.1 
contains a background discussion and Subsection 1.2 states our results.
Regarding quantum black-box algorithms, Subsection 1.3 
contains a background discussion and Subsection 1.4 states our results.
The results are all proven in Sections 2 and 3.

\subsection{Quantum communication complexity}

The recent book by Kushilevitz and Nisan \cite{KN96} is an excellent text 
on communication complexity.
As usual, two parties, Alice and Bob, wish to compute a boolean function
on their $N$-bit inputs using a communication protocol.
It will be convenient to let $N = 2^n$ and think of Alice and Bob's 
$N$-bit inputs as functions $f, g : \01^n \rightarrow \01$ 
(e.g., when $n=2$, $f$ represents the four-bit string 
$f(00)f(01)f(10)f(11)$).
Examples of well-studied communication problems are: 
\begin{itemize} 
\item Equality: 
$\EQ(f,g) = \bigwedge_{x \in \01^n} (f(x) = g(x))$
\item Inner product: 
$\IP(f,g) = \bigoplus_{x \in \01^n} (f(x) \wedge g(x))$
\item Disjointness%
\footnote{In fact this defines the {\em complement} of the set disjointness 
problem. Since for the models we study the communication complexity of 
$\DISJ$ and its complement are equal our results hold for both.}: 
$\DISJ(f,g) = \bigvee_{x \in \01^n} (f(x) \wedge g(x))$.
\end{itemize}

Classical communication  protocols were defined by Yao \cite{Ya79}. 
In an $m$-bit deterministic protocol, the players exchange $m$ 
(classical) bits according to their individual inputs and then decide on 
an answer, which must be correct.
In an $m$-bit probabilistic protocol, the players are allowed to flip 
coins to decide their moves, but they still must exchange at most $m$ 
bits in any run.
The answer becomes a random variable, and we demand that the answer be 
correct with probability at least $1 - \e$ (for some $\e \ge 0$) for every 
input pair.
Note that if $\e$ is set to 0 then probabilistic protocols are not more 
powerful than deterministic ones.

An alternative measure of the communication cost of a probabilistic 
protocol is to take the {\it expected\/} communication cost of a run,  
with respect to the outcomes of the coin flips (rather than the 
worst-case communication cost of a run).
In this case, probabilistic protocols with error probability zero may 
be more powerful than deterministic protocols.
Another alternate definition for probabilistic protocols is where the 
players share a random string.
This model has been shown to have the same power as the above 
bounded-error model whenever the communication complexity is above 
$\log N$ \cite{Ne91}.

For a communication problem $P$, and $\e \ge 0$, let $C_{\e}(P)$ denote 
the minimum $m$ such that there is a (probabilistic) protocol that requires 
at most $m$ bits of communication and determines the correct answer with 
probability at least $1-\e$.
Then $C_0(P)$ can be taken as the {\em deterministic} communication 
complexity of $P$ (sometimes denoted as $D(P)$).
Also, let $C(P)$ denote $C_{1/3}(P)$, the {\em bounded-error} communication 
complexity of $P$.
Clearly, $C(P) \leq C_0(P)$, and there are instances where there are 
exponential gaps between them.
Furthermore, let $C^E_0(P)$ denote the minimum {\em expected\/} communication 
for probabilistic errorless protocols, frequently called the {\em zero-error} 
communication complexity.
According to our definitions, ${2 \over 3}C(P) \leq C^E_0(P) \leq C_0(P)$.


For the aforementioned problems, the following is known.

\begin{theorem}\cite{Ya79}
$C_0(\EQ) = C^E_0(\EQ) = N$, but $C(\EQ) \in O(\log N)$.
\end{theorem}

\begin{theorem}\cite{CG88}
$C(\IP) \in \Omega(N)$.
\end{theorem}

\begin{theorem}\label{dis_lb} \cite{KS87,Raz90}
$C(\DISJ) \in \Omega(N)$.
\end{theorem}

Yao \cite{Ya93} also introduced a {\em quantum} communication complexity 
model, where Alice and Bob are allowed to communicate with qubits rather 
than bits.
It is not immediately clear whether using qubits can reduce communication 
because a fundamental result in quantum information theory by Holevo 
\cite{Hol73} (see also \cite{FC94}) implies that by sending 
$m$ qubits one cannot convey more than $m$ classical bits of information.
Yao's motivation was to prove lower bounds on the size of particular kinds 
of quantum circuits that compute the $\MAJ$ function, and he 
accomplished this via a qubit communication complexity lower bound.
The MSc thesis of Kremer \cite{Kr95} includes several important definitions 
and basic results.

Denote by $Q_{\e}(P)$ the minimum $m$ for which there is a protocol for 
$P$ involving $m$ qubits of communication with error probability bounded 
by $\e$.
Let $Q(P)$ denote $Q_{1/3}(P)$, the {\em bounded-error} communication 
complexity of $P$.
Also, call $Q_0(P)$ the {\em exact} communication complexity of $P$.
It turns out that one of the differences between the quantum scenario 
and the classical probabilistic scenario is that $Q_0(P)$ is not the 
same as the deterministic communication complexity of $P$ (see 
Theorem~\ref{bal} below), whereas $C_0(P)$ is.

A basic result is that quantum protocols are at least as powerful as 
probabilistic ones.

\begin{fact}\cite{Kr95} For every problem $P$ on $n$-bit inputs,
$Q(P) \leq C(P)$ and $Q_0(P) \leq C_0(P)$.
\end{fact}

\ni Kremer also presents the following lower bound (whose origin he 
attributes to Yao).

\begin{theorem}\cite{Kr95} (see also \cite{CDNT97}) \label{thm:lowerboundip}
$Q(\IP) \in \Omega(N)$.
\end{theorem}

\ni Kremer leaves open the question of whether the quantum (qubit) model 
is ever more powerful than the classical bit model for any communication 
problem.

Cleve and Buhrman \cite{CB97} (see also \cite{BCD97}) showed the first example 
where quantum information reduces communication complexity.
They considered a different model than that of Yao, the {\em entanglement\/} 
model, where the communication is restricted to classical bits; however, the 
parties have an {\em a priori} set of qubits in an entangled quantum state.
As with the qubit model, there are no trivial communication advantages 
in the entanglement model, because a prior entanglement cannot reduce 
the communication cost of conveying $m$ bits.
In this model, they demonstrated a three-party communication problem where 
the prior entanglement reduces the required communication complexity by 
one bit.
Buhrman \cite{Bu97} showed that, in this model, the separation between 
quantum vs.\ classical communication costs can be as large as 
$2n$ vs.\ $3n$.
Also, van Dam, H\o yer, and Tapp \cite{DHT97} showed the first instance 
where the reduction in communication can be asymptotically large in a 
multi-party setting.
They showed that, for a particular $k$-party scenario, the quantum vs.\ 
classical communication cost is roughly $k$ vs.\ $k \log k$ 
(note that this falls short of an asymptotic separation when the number 
of parties is fixed).

\subsection{Our results in quantum communication complexity}

We prove some asymptotic gap theorems between quantum and classical 
two-party communication.
The first is a near quadratic gap for the bounded-error models 
(and also happens to be a near quadratic gap between $Q$ and $Q_0$).

\begin{theorem}\label{dis}
$Q(\DISJ) \in O(\sqrt N \log N )$ and \\ $ Q_0(\DISJ) \in \Omega(N)$.
\end{theorem}

This, combined with Theorem \ref{dis_lb}, results in a near quadratic 
separation between classical bounded-error communication complexity 
and quantum bounded-error communication complexity.

Our second theorem is an exponential gap between the exact quantum and 
the zero-error classical model.  For this, we need to define a 
{\em partial\/} function.
Let $\Delta(f,g)$ denote the hamming distance between the two functions $f,g$
(viewed as binary strings of length $N=2^n$).  Define the partial function
$\EQprime$ as $$\EQprime(f,g) = \cases{1 & if $\Delta(f,g)=0$\cr 0 & if
$\Delta(f,g)=2^{n-1}$.\cr} $$ For a partial function all communication
definitions above extend in the natural way, demanding correct (or 
approximately correct) answers only for pairs on which the partial function 
is defined.

\begin{theorem}\label{bal}
$Q_0 (\EQprime) \in O(\log N)$, but 
$C_0(\EQprime), C^E_0(\EQprime) \in \Omega(N)$.
\end{theorem} 

Finally, we generalize Theorem \ref{dis} to balanced, constant depth
formulae.

\begin{theorem}\label{AC0}
Let $F$ be any balanced depth-$d$ $AC^0$ formula (i.e. formula with 
unbounded fan-in $\wedge$ and $\vee$ gates) with $N$ leaves, and 
$L : \01^2 \rightarrow \01$.
Then the communication problem $P(f,g) = F(L(f,g))$ 
has complexity $O(\sqrt{N} \log^{d-1}(N))$.
\end{theorem}

The classical lower bounds will appeal to known techniques and results in
communication complexity and combinatorics.
The quantum upper bounds will follow from a reduction from communication 
problems to computational problems where the input is given as a black-box, 
in conjunction with known quantum algorithms for these problems---and 
a new quantum algorithm in the case of Theorem \ref{AC0}.
The reduction is presented in Theorem~\ref{thm:simulation}, 
Section~\ref{reduction}. Applying this reduction in its reverse direction 
enables us to translate lower bounds for quantum communication problems 
into lower bounds for black-box computations.

\subsection{Black-box quantum computations}

All the upper bounds on communication complexity will come from a 
simulation of quantum circuits whose inputs are functions that 
can be queried as black-boxes.
Relevant definitions (and some lower bound techniques) may be 
found in \cite{BV93,BB94,BBBV93,Ya93}.

For $f : \01^n \rightarrow \01$, define an {\em $f$-gate} as the 
unitary mapping such that
\begin{equation}
U_f : \ket{x}\ket{y} \mapsto \ket{x}\ket{f(x) \oplus y},
\label{gate}
\end{equation}
for all $x \in \01^n$, $y \in \01$.
For the initial state with $x \in \01^n$ and $y=0$, this mapping simply 
writes the value of $f(x)$ on the $n+1^{\mbox{\scriptsize st}}$ qubit; 
however, for this gate to make sense when evaluated in quantum 
superposition, it must also be defined for the $y=1$ case as well as be 
reversible.

A {\em quantum circuit} (or {\em gate array}) $G$ with input given as a 
black-box operates as follows.
It begins with a set of qubits in some initial state (say, 
$\ket{0,\ldots,0}$) and performs a sequence of unitary transformations 
to this state.
These unitary transformations are from a designated set of ``basis'' 
operations (say, the set of all operations corresponding to ``two-qubit 
gates''), as well as $f$-gates.
At the end of the computation, the state is measured (in the {\em standard\/} 
basis, consisting of states of the form $\ket{x_1,\ldots,x_m}$, for 
$x_1,\ldots,x_m \in \01$), and some designated bit (or set of bits) is taken 
as the {\em output\/}.
Denote the output of $G$ on {\em input\/} $f$ as $G(f)$ 
(which is a random variable).

Let $\H$ be a collection of functions.
We say that a quantum circuit $G$ {\em computes} a function 
$F: \H \ra S$ {\em with error} $\varepsilon$ if, for every $h\in \H$, 
$\Pr[G(h) = F(h)] \geq 1 - \varepsilon$.
We denote by $T_{\varepsilon}(F)$ the minimum $t$ (time, or, more accurately, 
number of black-box accesses) for which there is a quantum circuit that 
computes $F$ with error $\varepsilon$.
We call $T_0(F)$ the {\em exact} quantum complexity of $F$, and 
we call $T(F) = T_{1/3}(F)$ the {\em bounded-error} quantum complexity of 
$F$.

Here are three well-known examples of nontrivial quantum algorithms (and 
precious few others are known).
For these problems, classical (probabilistic) computations require 
$\Theta(2^n)$, $\Theta(\sqrt{2^n})$, and $\Theta(2^n)$ black-box queries 
(respectively) to achieve the same error probability as the quantum 
algorithm.

\begin{itemize}

\item
{\bf Half or None} Here $\H$ consists of 
the constant $0$ function and all ``balanced'' functions
(i.e.\ $h$'s which take on an equal number of $0$s and $1$s).
The function $\BAL$ takes the value $1$ if $h$ is balanced and $0$
otherwise.
\begin{theorem}\cite{DJ92}\label{DJ}
$T_0(\BAL) = 1$.
\end{theorem}

\item
{\bf Abelian Subgroups} Here $\H$ are all functions 
$h : \01^n \rightarrow \01$ for which there exists a subgroup $K$ of $Z_2^n$ 
(represented by $\01^n$) such that $h(x) = h(y)$ iff $x+y \in K$.
$\STAB(h)$ is a specification of $K$.
\begin{theorem}\cite{Si94}\label{Si94}
$T(\STAB) \in O(n)$.
\end{theorem}
This theorem has been generalized (appropriately) to other Abelian 
groups by Kitaev \cite{Ki95}.

\item
{\bf Database Search}
Here $\H$ contains {\em all} possible functions $h : \01^n \rightarrow \01$ 
and 
$$\OR(f) = \bigvee_{x \in \01^n} f(x).$$
Based on a technique introduced by Grover and later refined by Boyer 
{\it et al.}, it is straightforward to construct a quantum 
algorithm that solves $\OR$ with the following efficiency.
\begin{theorem}\cite{Gr96,BBHT96}\label{Gr96}
$T(\OR) \in O(\sqrt{2^n})$.
\end{theorem}

\end{itemize}

\subsection{Our results about black-box quantum computations}

Define 
$$\PARITY(f) = \bigoplus_{x \in \01^n} f(x).$$
$\PARITY$ is at least as hard as $\OR$ (in the bounded-error 
case) by a result of Valiant and Vazirani \cite{VV86}.
In view of Theorem \ref{Gr96}, it is natural to ask whether 
Grover's technique can somehow be adapted to solve $\PARITY$ with 
quadratic speedup---or at least to solve $\PARITY$ in $O((2^n)^r)$ steps 
for some $r<1$.
We show that this is not possible by the following.

\begin{theorem}\label{parity}
$T(\PARITY) \in \Omega(2^n/n)$.
\end{theorem}

\ni Also, define 
$$
\MAJ(f) = 
\cases{1 & if $\sum_{x \in \01^n} f(x) > 2^{n-1}$\cr
       0 & otherwise.\cr}
$$

\begin{theorem}\label{majority}
$T(\MAJ) \in \Omega(2^n/n^2\log(n))$.
\end{theorem}

\ni Considering this dichotomy among Theorems \ref{Gr96} to \ref{majority}, 
we investigate the bounded-depth predicates 
(i.e.\ the polynomial-time hierarchy).
First, define $\SIGMA_2$, on functions 
$f : \01^{n} \rightarrow \01$ as 
\begin{eqnarray}
\lefteqn{\SIGMA_2(f) =} & & \nonumber \\
& & \bigvee_{x \in \01^{n-m}} 
\left(\bigwedge_{y \in \01^{m}} f(x_1,\ldots,x_{n-m},y_1,\ldots,y_m)\right) 
\ \ \ \ \ \ 
\end{eqnarray}
(where $m \in \{0,\ldots,n\}$ is an implicit parameter of $\SIGMA_2$).
Let $\PI_2$ be the negation of $\SIGMA_2$.
Both $\PI_2$ and $\SIGMA_2$ have  classical complexity $\Theta(2^{n})$.
Is a near square root speed up possible for these problems?
We shall show 

\begin{theorem}\label{sigma2}
$T(\SIGMA_2) \in O(\sqrt{2^n}\,n)$.
\end{theorem}

\ni More generally, for the appropriate generalization to higher 
levels of the polynomial-hierarchy, define for $d$ alternations
\begin{eqnarray}
\lefteqn{\SIGMA_d(f) =} & & \nonumber \\
& & \bigvee_{x^{(1)} \in \01^{m_1}} 
\left( 
\cdots
\left(\bigwedge_{x^{(d)} \in \01^{m_d}}
f(x^{(1)},\ldots,x^{(d)})
\right) 
\cdots
\right) \ \ \ \ \ \ 
\end{eqnarray}
(where $m_1,\ldots,m_d$ are implicit parameters with 
$m_1 + m_2 + \cdots + m_d = n$), and $\PI_d$ is the negation of $\SIGMA_d$.
Note that $\SIGMA_1 = \OR$ and $\PI_1$ is equivalent to $\AND$.
\begin{theorem}\label{ph}
$T(\SIGMA_d) \in O(\sqrt{2^{n}}\,n^{d-1})$ and \\
$T(\PI_d) \in O(\sqrt{2^{n}}\,n^{d-1})$.
\end{theorem}

\ni This is a near square root speed up for any fixed value of $d$.

Moreover, if we are willing to settle for speed up by a root slightly worse 
that square, such as $O((2^n)^{1/2 + \delta})$ steps (for some fixed 
$\delta > 0$), then the error probability can be {\em double exponentially} 
small!

\begin{theorem}\label{double_exp}
For $\e = 1 / 2^{2^{{}^{(n/ \delta d)-1}}}$, 
$T_{\e}(\SIGMA_d) \in O((2^n)^{1/2 + \delta})$.
\end{theorem}

\ni Finally, in sharp contrast with Theorem~\ref{Gr96} and 
Theorem~\ref{double_exp}, we have 

\begin{theorem}\label{errorless}
$T_0(\OR) \in \Omega(2^n/n)$.
\end{theorem}

\section{Reducing communication to computation 
problems}\label{reduction}

In this section, we prove a central theorem of this paper, which is 
essentially a simulation technique that transforms quantum algorithms 
for black-box computation to quantum communication protocols.
While the idea of the simulation is extremely simple, we stress that
it utilizes quantum parallelism in full.

This enables us to obtain new quantum communication protocols by 
applying the simulation to known quantum algorithms.
We can also apply this technique in the reverse direction to 
use lower bounds for quantum communication protocols to derive 
lower bounds for quantum computation.

Let $\F_n$ denote the set of all functions $f : \01^n \rightarrow \01$.

\begin{theorem}\label{thm:simulation}
Let $F : \F_n \rightarrow \01$ and 
$L : \01 \x \01 \rightarrow \01$.
$L$ induces a mapping $\F_n \x \F _n\rightarrow \F_n$ by 
pointwise application: $L(g,h)(x) = L(g(x),h(x))$, for all $x \in \01^n$.
If there is a quantum algorithm that computes $F(f)$ for input $f$ 
using $t$ $f$-gate calls then there is a $t(2n+4)$ qubit communication 
protocol for the following problem.
Alice gets $g$, Bob gets $h$ and the goal is for Alice to determine 
$F(L(g,h))$.
Furthermore, if the algorithm succeeds with a certain probability 
then the corresponding protocol succeeds with the same probability.
\end{theorem}

\prf
Consider the quantum circuit $G$ that computes $F(f)$, with $t$ $f$-gate 
calls.
In the communication protocol, Alice simulates the quantum circuit $G$ 
with $f$ set to $L(g,h)$.
She communicates with Bob only when an $L(g,h)$-gate call is made 
(for which she needs Bob's help, since she does not know $h$).
Note that Alice has sufficient information to simulate a $g$-gate 
and Bob has enough information to simulate an $h$-gate.
Each $L(g,h)$-gate call is simulated by the following procedure 
for the state $\ket{x}\ket{y}$ (for each $x \in \01^n$ and $y \in \01$).
\begin{enumerate}
\item Alice sets an ``ancilla'' qubit to state $\ket{0}$.
\item Alice applies the mapping 
$\ket{x}\ket{y}\ket{0} \mapsto 
\ket{x}\ket{y}\ket{g(x)}$, and then sends the $n+2$ qubits to Bob.
\item Bob applies 
$\ket{x}\ket{y}\ket{g(x)} \mapsto 
\ket{x}\ket{L(g(x),h(x)) \oplus y}\ket{g(x)}$, 
and then sends the $n+2$ qubits back to Alice.
\item Alice applies
$\ket{x}\ket{L(g(x),h(x)) \oplus y}\ket{g(x)} \mapsto$\newline
$\ket{x}\ket{L(g(x),h(x)) \oplus y}\ket{0}$.
\end{enumerate}
This involves $2n + 4$ qubits of communication.
Therefore, the total amount of communication is $t(2n+4)$ qubits.
\qed

Note that it is simple to generalize Theorem~\ref{thm:simulation} to
functions whose range is an arbitrary set $S$ instead of $\{0,1\}$,
and any $L:S\times S \rightarrow \{0,1\}$.

\section{Proofs of upper and lower bounds}

\noindent
{\bf Proof of Theorem \ref{dis}}

The upper bounds follows directly from the simulation result Theorem 
\ref{thm:simulation} with $L$ being the binary $\AND$ function and $F$ the 
$2^n$-ary $\OR$ function, together with the quantum algorithm for $\OR$ 
referred to in Theorem~\ref{Gr96}.

The lower bound on $Q(\DISJ)$ is the well known result of
Kalyanasundaram and Schnitger \cite{KS87} (see also \cite{Raz90}), stated
in Theorem \ref{dis_lb}.

It remains to prove the linear lower bound on $Q_0$.
By results in \cite{Kr95}, it is straightforward to see that 
see  that a zero-error $m$-qubit quantum protocol for a communication 
problem $P$ puts an upper bound of $2^{O(m)}$ on the rank (over the reals) 
of the matrix describing $P$ (more details will be provided in the final 
version).
It is well known that the set disjointness matrix has full rank over 
the reals, which gives $m = \Omega(n)$.\ee

\noindent
{\bf Proof of Theorem \ref{bal}}

The upper bound follows directly from the simulation result Theorem
\ref{thm:simulation} with $L$ being the binary XOR function and $F$ the
$N$-way $\OR$ function (restricted to balanced or zero inputs), 
together with the Deutsch-Jozsa algorithm for $F$
\cite{DJ92} stated in Theorem~\ref{DJ}.
We observe that the algorithm accesses the $F$-gate only once, and in fact 
a {\em 1-way} communication protocol of $O(\log N)$ qubits can be obtained 
in this case.

For the lower bound we need the following strong result of Frankl and
R\"odl \cite{FR87}, which seems tailor-made for our needs.

\begin{theorem}\label{FR}\cite{FR87}
Let $n$ be divisible by 4.
Let $S,T \subseteq \zon$ be two families of $n$-bit vectors, such that
for every pair $s\in S, t\in T$ we have $\Delta(s,t) \neq n/2$.
Then $|S|\times|T| \leq 4^{.96n}$.
\end{theorem}

Consider any deterministic protocol for $\EQprime$ 
of complexity less than $n/100$.
Take the largest rectangle answering 1, and let its 
sides be the subsets $S,T$. Since there are $2^n$ 1 answers, 
and they lie on the diagonal, we must 
have $|S|\times|T| \geq 4^{.98n}$. On the other hand, since this 
algorithm makes no error, this rectangle has no 0 entry, which means 
that for every pair $s\in S, t\in T$ we have $\Delta(s,t) \neq n/2$, 
which is a contradiction.

For the probabilistic, zero-error lower bound, 
it suffices to give a distribution which is hard on average for
deterministic protocols.
Observe that the argument
above shows that the depth of {\em every} 1- leaf of any protocol which
is always correct, is at least $n/100$. By considering an input
distribution which places half the weight on each of the two output
values, and then uniformly on the input pairs for each output,
requires average communication $n/100 -1$ in every such protocol.\ee

\noindent
{\bf Proof of Theorem \ref{AC0}}

The upper bound follows directly from the simulation result Theorem
\ref{thm:simulation} with $L$ being the binary $\AND$ function and $F$ 
the $AC^0$ formula corresponding a $\SIGMA_d$ or $\PI_d$ 
predicate, together with Theorem~\ref{ph}.\ee

\noindent
{\bf Proof of Theorem \ref{parity}}

Let $F(f)$ be $\PARITY(f)$. Suppose we have a quantum algorithm for
$F$ that makes $t$ $f$-gate calls. Apply Theorem~\ref{thm:simulation}.
Let $L$ be the $\AND$ function and observe that $F(L(g,h))$ is the inner 
product communication problem.  An application of the lower bound for 
$Q(\IP)$, Theorem~\ref{thm:lowerboundip} yields that $t \in \Omega(2^n/n)$.\ee

\noindent
{\bf Proof of Theorem \ref{majority}}

The problem of computing $\PARITY$ with error $1 \over 3$ can be reduced 
to $n$ instances of computing $\MAJ$ with error $O({1 \over n})$.
The latter problem reduces to $O(\log n)$ instances of computing $\MAJ$ 
with error $1 \over 3$.
Therefore, in the bounded-error model, $\PARITY$ is reducible to 
$n \log n$ instances of $\MAJ$.
The result now follows from Theorem \ref{parity}.\ee

\noindent
{\bf Proof of Theorem \ref{sigma2}}

The basic approach is to define
\begin{equation}
g(x_1,\ldots,x_{n-m}) = 
\bigwedge_{y \in \01^m} f(x_1,\ldots,x_{n-m},y_1,\ldots,y_m)
\end{equation}
and then first use Boyer {\it et al.}'s \cite{BBHT96} extension of 
Grover's technique \cite{Gr96} (in a way that does not involve measurements) 
to simulate an {\em approximate} $g$-gate within accuracy 
$\varepsilon / \sqrt{2^{n-m}}$.
More precisely, a $g$-gate is a unitary transformation
\begin{equation}
U_g : \ket{x_1,\ldots,x_{n-m}}\ket{z} \mapsto 
\ket{x_1,\ldots,x_{n-m}}\ket{z \oplus g(x)}, 
\end{equation}
and we'll simulate a unitary transformation $V$ such that 
$\| U_g - V \|_2 \le \varepsilon / \sqrt{2^{n-m}}$ (where $\| \cdot \|_2$ 
is the norm induced by Euclidean distance).
We'll see that this can be accomplished unitarily with $O(\sqrt{2^m}\,n)$ 
accesses to the $f$-gate.

Then the Grover technique is applied to compute 
\begin{equation}
\bigvee_{x \in \01^{n-m}} g(x_1,\ldots,x_{n-m})
\end{equation}
 with $O(\sqrt{2^{n-m}})$ 
calls to the $g$-gate.
Due to the accuracy of our simulated approximate $g$-gate calls, 
they can be used in place of the true $g$-gate calls, and the resulting 
total accumulated error in the final state will be bounded by $\varepsilon$.
This follows from the unitarity of the operations (see \cite{BBBV93}).
This inaccuracy affects the correctness probability of the final measured 
answer by at most $2 \varepsilon$.

It remains to show how to compute the approximate $g$-gates.
In \cite{BBHT96}, it is shown that the Grover search procedure can be 
implemented so as to find a satisfying assignment (whenever one exists) 
of an $m$-variable function with an expected number of $O(\sqrt{2^m})$ 
calls to that function (and this holds without knowing anything 
about the number of satisfying assignments).
Their procedure essentially involves a sequence of independent runs of 
Grover's original procedure for various carefully chosen run lengths.
By stopping this after an appropriate number of runs, we obtain a procedure 
that, with $c \sqrt{2^m}$ black-box calls, decides the satisfiability of 
the function with error probability at most $\half$ (and only errs in 
the case of satisfiability).
By repeating this $k$ times, we obtain a procedure that, with $c k \sqrt{2^m}$ 
queries, decides the satisfiability of the function with error 
probability at most $2^{-k}$.
This procedure will involve several intermediate measurements; however, 
by standard quantum computing techniques, the procedure can be modified 
so that it runs for a purely unitary stage, $G$, followed by a single 
measurement step.

In our context,  $G$ can be thought of as being applied on 
an initial quantum state of the form 
$\ket{x_1,\ldots,x_{n-m}}\ket{0,\ldots,0}$, 
(for some $x_1,\ldots,x_{n-m} \in \01$) and making calls to $f$-gates, 
with the first $n-m$ inputs of $f$ always set to state 
$\ket{x_1,\ldots,x_{n-m}}$.
What we know about the state after applying $G$ is that, if its first qubit 
(say) is measured, the result will be $g(x)$ with probability at least 
$1 - 2^{-k}$.
This means that, after applying $G$ to 
$\ket{x_1,\ldots,x_{n-m}}\ket{0,\ldots,0}$, 
but prior to any measurements, the state must be of the form
\begin{equation}
\alpha \ket{g(x)}\ket{A} + \beta \ket{\overline{g(x)}}\ket{B},
\end{equation}
where $|\alpha|^2 \ge 1-2^{-k}$ and $|\beta|^2 \le 2^{-k}$.

Now, consider the following construction.
Introduce a new qubit, in initial state $\ket{z}$ (for some $z \in \01$) 
and apply the following steps to the state 
$\ket{z}\ket{x_1,\ldots,x_{n-m}}\ket{0,\ldots,0}$:
\begin{enumerate}
\item
Apply $G$.
\item
Perform a controlled-NOT with the first qubit as target and the second 
qubit as control (recall that here the second qubit contains the ``answer'' 
$g(x)$).
\item
Apply $G^{\dagger}$.
\end{enumerate}
(We'll show that this approximates the $g$-gate.)
Let us trace through the evolution of a {\em basis} state
$\ket{z}\ket{x_1,\ldots,x_{n-m}}\ket{0,\ldots,0}$.
After the $G$ operation, the state is 
\begin{equation}
\ket{z}\left(\alpha \ket{g(x)}\ket{A} 
+ \beta \ket{\overline{g(x)}}\ket{B}\right).
\end{equation}
After the controlled-NOT gate, the state is
\begin{eqnarray}
\lefteqn{\alpha \ket{z \oplus g(x)}\ket{g(x)}\ket{A} 
+ \beta \ket{\overline{z \oplus g(x)}}\ket{\overline{g(x)}}\ket{B}} & & 
\nonumber \\
& = & \alpha \ket{z \oplus g(x)}\ket{g(x)}\ket{A} 
+ \beta \ket{z \oplus g(x)}\ket{\overline{g(x)}}\ket{B} - \nonumber  \\
 & & \beta \ket{z \oplus g(x)}\ket{\overline{g(x)}}\ket{B}
+ \beta \ket{\overline{z \oplus g(x)}}\ket{\overline{g(x)}}\ket{B}
\nonumber \\
& = & \ket{z \oplus g(x)}
\left(\alpha \ket{g(x)}\ket{A} + \beta \ket{\overline{g(x)}}\ket{B}\right) +\\
\nonumber & & \sqrt{2}\beta
\left(\sqhalf\ket{\overline{z \oplus g(x)}} - \sqhalf\ket{z \oplus g(x)}\right)
\ket{\overline{g(x)}}\ket{B}.
\end{eqnarray}
In this form, it's easy to see that, after applying $G^{\dagger}$, the 
state is 
\begin{equation}
\ket{z \oplus g(x)}\ket{x_1,\ldots,x_{n-m}}\ket{0,\ldots,0} + 
\end{equation}
\[\sqrt{2}\beta
\left(\sqhalf\ket{\overline{z \oplus g(x)}} - \sqhalf\ket{z \oplus g(x)}\right)
G^{\dagger}\ket{\overline{g(x)}}\ket{B}.
\]
The Euclidean distance between this state and the state that a true 
$g$-gate would produce is 
$\sqrt{2}|\beta| \le \sqrt{2} \cdot 2^{-k/2}$.
The above distance holds for any initial basis state 
$\ket{z}\ket{x_1,\ldots,x_{n-m}}\ket{0,\ldots,0}$; however, the distance 
might be larger for non-basis states.
In general, the input to a $g$-gate is of the form 
\begin{equation}
\sum_{z \in \01 \atop x \in \01^{n-m}} 
\lambda_{z,x} \ket{z}\ket{x_1,\ldots,x_{n-m}}\ket{0,\ldots,0}, 
\end{equation}
where $\sum_{z \in \01, x \in \01^{n-m}} |\lambda_{z,x}|^2 = 1$.
In this case, the difference between the output state of the true 
$g$-gate and our approximation to it is still bounded by 
\[
\sum_{z \in \01 \atop x \in \01^{n-m}} |\lambda_{z,x}| \sqrt{2} \cdot 2^{-k/2} 
\le \sqrt{2^{n-m+1}} \cdot \sqrt{2} \cdot 2^{-k/2}\]
\begin{equation}
 = 2^{{n-m \over 2} + 1 - {k \over 2}}.
\end{equation}

Now, in order to make this quantity bounded by $\varepsilon / \sqrt{2^{n-m}}$, 
it suffices to set $k \ge 2(n-m) + 2 \log(2/\varepsilon)$.

Thus, the total number of $f$-gate calls is 
$O(\sqrt{2^{n-m}} \cdot 2 \cdot c \cdot (2(n-m) + 2 \log(2/\varepsilon)) 
\cdot \sqrt{2^m}) \subseteq O(\sqrt{2^n}\,n)$, as claimed.\ee

\noindent
{\bf Proof of Theorem \ref{ph}}
This is a straightforward generalization of Theorem \ref{sigma2}.
For each $i \in \{1,\ldots,d\}$, define 
\begin{eqnarray}
\lefteqn{g^{(i)}(x^{(1)},x^{(2)},\ldots,x^{(i-1)}) =} & & \nonumber \\
& & \bigvee_{x^{(i)} \in \01^{m_i}}
\cdots \left(\bigwedge_{x^{(d)} \in \01^{m_d}} f(x^{(1)},\ldots,x^{(d)})
\right) \ \ \ \ 
\end{eqnarray}
(where the $\wedge$ and $\vee$ quantifiers are appropriately placed).
As in the proof of Theorem \ref{sigma2}, an approximation of 
$g^{(d)}$ is first constructed at a cost of $c k_d \sqrt{2^{m_d}}$.
Then this is used to approximate $g^{(d-1)}$ with cost 
$(c k_{d-1} \sqrt{2^{m_{d-1}}})(c k_d \sqrt{2^{m_d}})$, and so on, 
up to $g^{(1)}$, whose value is the required answer.
It suffices to set $k_2,\ldots,k_{d}$ to $5n$ and to set $k_1$ to 
a constant.\ee

\ni {\bf Proof of Theorem \ref{double_exp}}

This is similar to the proof of Theorem \ref{ph}, except that 
the parameters $k_1,\ldots,k_d$ are all set to $2^{n / \delta d}$.\ee

\ni {\bf Proof of Theorem \ref{AC0}}

This follows from the zero-error part of Theorem \ref{dis} in conjunction 
with Theorem \ref{thm:simulation}.

\section{Conclusions and open problems}
We have constructed a reductions from quantum communication problems to quantum
black-box computations. Using known quantum algorithms, this reduction enabled
us to prove a near quadratic gap between bounded error classical communication
complexity and bounded error quantum communication complexity. Using a partial
function we also showed  an exponential gap between zero-error classical
communication complexity and exact quantum communication
complexity. Kremer~\cite{Kr95} shows that the gap between the two models can
never be bigger than exponential, so this result is optimal. Several problems
however remain:
\begin{itemize}
\item Is there an exponential gap between the exact and the zero-error model
with a total instead of a partial function? A recent result by \cite{BBCMW98}
shows that for any total black-box problem if there is a quantum algorithm that
computes this problem with $T$ oracle calls then there is a deterministic
classical algorithm that computes it with $O(T^6)$ oracle calls. This results
shows that the approach taken here (reduce a communication  problem
to a black-box problem) will for total functions never yield more than a
polynomial (sixth root) gap.
\item Is the upper bound for $\DISJ$ optimal?
\item Is there a bigger than quadratic  gap for the bounded-error models 
(with total or partial functions)? 
\end{itemize}

We used the reduction from communication  problems to black-box
computation in the reverse order to obtain non-trivial lower bounds for
$\PARITY$ and $\MAJ$. These bounds have recently been improved to optimal for
$\PARITY$~\cite{BBCMW98,FGGS98} and $\MAJ$~\cite{BBCMW98}.

\section*{Acknowledgements}
We would like to thank Gilles Brassard, Wim van Dam, Peter H\o yer, Tal Mor, 
and Alain Tapp for helpful discussions.
A.W. would like to thank Dorit Aharonov, who taught him almost all he 
knows about quantum computing. This is the ultimate student-advisor 
relationship!

\end{document}